\documentclass[superscriptaddress,altaffilletter,amsmath,amssymb,aps,twocolumn,floatfix]{revtex4-1}
\pdfoutput=1
\usepackage[english]{babel}
\usepackage[colorinlistoftodos]{todonotes}
\usepackage{multirow}
\usepackage{comment}
\usepackage{relsize}
\usepackage{float}
\usepackage{textcomp}
\usepackage{hyperref}
\usepackage[justification=justified]{subcaption}
\usepackage[justification=justified, figurename=FIG.]{caption}
\usepackage[compact]{titlesec}
\titlespacing{\section}{0pt}{3ex}{2ex}
\titlespacing{\subsection}{0pt}{2ex}{2ex}
\titlespacing{\subsubsection}{0pt}{1.5ex}{1.5ex}

\DeclareCaptionLabelFormat{UC}{\MakeUppercase{#1} #2}
\usepackage{xcolor}
\usepackage{graphicx}
\usepackage{epstopdf}
\usepackage{color}
\usepackage{dcolumn}
\usepackage{bm}
\usepackage{hyperref}
\usepackage[mathlines]{lineno}
\usepackage{threeparttable}
\usepackage{booktabs} 
\usepackage{siunitx}
\usepackage{placeins}
\usepackage{xspace}

\newcommand{\dbd}{\ensuremath{0\nu\beta\beta}\xspace} 
\newcommand{\tnbd}{\ensuremath{2\nu\beta\beta}\xspace} 
\newcommand{\gaeff}{\ensuremath{g_\textrm{A}^\textrm{eff}}\xspace} 
\newcommand{\XeoTo}{\ensuremath{^{131}\textrm{Xe}}\xspace} 
\newcommand{\XeoTs}{\ensuremath{^{136}\textrm{Xe}}\xspace} 
\newcommand{\CsoTo}{\ensuremath{^{131}\textrm{Cs}}\xspace} 
\newcommand{\CsoTs}{\ensuremath{^{136}\textrm{Cs}}\xspace} 

\newcommand\Tstrut{\rule{0pt}{2.6ex}}         
\newcommand\Bstrut{\rule[-0.9ex]{0pt}{0pt}}   

\begin{document}

\title{Solar neutrino detection in liquid xenon detectors via charged-current scattering to excited states}

\author{Scott Haselschwardt} \email[]{scotthaselschwardt@lbl.gov}
\affiliation{Lawrence Berkeley National Laboratory, 1 Cyclotron Road, Berkeley, CA 94720, USA}
\author{Brian Lenardo} \email[]{blenardo@stanford.edu}
\affiliation{Stanford University, Department of Physics, 382 Via Pueblo, Stanford, CA 94305, USA}
\author{Pekka Pirinen} \email{pekka.a.pirinen@jyu.fi}
\author{Jouni Suhonen}

\affiliation{University of Jyvaskyla, Department of Physics, P. O. Box 35 (YFL), FI-40014, Finland}

\date{\today}
\begin{abstract}
\noindent
We investigate the prospects for real-time detection of solar neutrinos via the charged-current neutrino-nucleus scattering process in liquid xenon time projection chambers. We use a nuclear shell model, benchmarked with experimental data, to calculate the cross sections for populating specific excited states of the caesium nuclei produced by neutrino capture on $^{131}$Xe and $^{136}$Xe. The shell model is further used to compute the decay schemes of the low-lying $1^{+}$ excited states of $^{136}$Cs, for which there is sparse experimental data. We explore the possibility of tagging the characteristic de-excitation $\gamma$-rays/conversion electrons using two techniques: spatial separation of their energy deposits using event topology and their time separation using delayed coincidence. The efficiencies in each case are evaluated within a range of realistic detector parameters. We find that the topological signatures are likely to be dominated by radon backgrounds, but that a delayed coincidence signature from long-lived states predicted in $^{136}$Cs may enable background-free detection of CNO neutrino interactions in next-generation experiments with smaller uncertainty than current measurements. We also estimate the sensitivity as a function of exposure for detecting the solar-temperature-induced line shift in $^{7}$Be neutrino emission, which may provide a new test of solar models.
\end{abstract}

\maketitle


\section{\label{sec:intro}INTRODUCTION}

Forthcoming experiments based on the liquid xenon (LXe) time projection chamber (TPC) plan to achieve new milestones in their overall sensitivity for rare processes. These include both experiments searching for direct scattering of dark matter and those pursuing an observation of neutrinoless double beta (\dbd) decay in \XeoTs. The current generation of dark matter experiments will deploy active targets containing $\sim5$ tonnes of natural xenon~\cite{LZ_WIMP_Sensitivity2019,XnT_sensitivity:2020}, and future \dbd decay experiments propose to use isotopically enriched targets of similar size~\cite{NEXO_pCDR:2018eqi}. The next-generation of dark matter experiments envision scaling this size by at least an order of magnitude~\cite{Aalbers:2016jon}. Detectors at this scale, typically designed to meet stringent low-background requirements, are of interest for their ability to measure rare interactions beyond their primary science goals.

A particularly interesting class of signals arise from the flux of neutrinos produced in the Sun, which simultaneously serve as an experimental probe into solar dynamics and fundamental neutrino physics~\cite{Dutta:2019oaj}. The study of solar neutrinos has been an active area of research for the last half-century. Experiments seeking to detect the low-energy ($\lesssim2$~MeV) components of the solar neutrino flux have thus far relied on one of two techniques: (1) after-the-fact chemical extraction of the isotope produced from neutrino charged-current (CC) capture on nuclei used in radiochemical experiments such as the Homestake~\cite{Cleveland:1998nv}, GALLEX/GNO~\cite{Kaether:2010ag}, and SAGE~\cite{Abdurashitov:1994bc} experiments, or (2) detecting the recoiling electron from neutrino-electron elastic scattering events in kiloton-scale liquid scintillator detectors such as Borexino~\cite{Alimonti:2008gc}. The former have the advantage of unambiguous detection of a neutrino interaction by directly detecting the resulting nucleus, but the disadvantage that only an integrated neutrino interaction rate above the reaction threshold is measured, and the contribution from different components must be inferred indirectly. The latter perform real-time spectroscopic measurements of neutrino interactions and can be sensitive to all low-energy components of the solar spectrum. However, the recoiling electron signal is indistinguishable from electrons generated by standard $\beta$-decay or Compton scattering, demanding strict control of background sources.

Though there has been a great deal of success using these techniques, it is interesting to consider combining the best features of both; that is, an experiment which can provide real-time spectroscopic information on the incoming neutrinos and can detect them via the CC capture process that allows the reduction of backgrounds through a tag of the resulting nucleus. Scattering events which populate excited states of the product nucleus can be identified through a tag of the isotope-specific $\gamma$-rays and/or internal conversion (IC) electrons. The presence of long lived states in the product nucleus can additionally provide very high suppression of background by way of the associated delayed coincidence signature. Over the last half-century, a variety of targets for tagged, low-energy solar neutrino capture have been proposed~\cite{Raghavan:1976yc,Raghavan:1997ad,Ejiri:1999rk,Zuber:2002wi,Wang:2020tdm}. The fact that $\beta\beta$-decay isotopes are particularly well suited for this reaction (often with a delayed coincidence) was pointed out by Raghavan~\cite{Raghavan:1997ad}. To date, none of these techniques have been realized at scale.

Here we explore this possibility in next-generation LXe detectors by studying the capture of solar neutrinos via the CC interaction on xenon nuclei: $\nu_e + \,^{A}_{54}\textrm{Xe} \to \,^{A}_{55}\textrm{Cs}^* + e^{-}$. Table~\ref{tab:isos} gives the corresponding reaction threshold for each naturally occurring isotope of xenon. In this paper we focus on interactions with \XeoTo and \XeoTs, the isotopes with the lowest thresholds. These low thresholds permit solar neutrinos from $^{7}$Be decay (primary energy 862~keV) and the carbon-nitrogen-oxygen (CNO) cycle (endpoint 1.7~MeV) to produce $\mathcal{O}(100\textrm{'s})$~keV primary electrons and populate low-lying excited states of the resulting Cs nuclei. The excellent position, energy, and timing information of modern LXe TPC experiments may allow for tagging of the emitted $\gamma$-rays and IC electrons, reducing backgrounds.

\begin{table}[t!]
\centering
\caption{Abundances~\cite{abund:2011} and thresholds for electron neutrino capture to the product Cs ground state of naturally occurring xenon isotopes. Only \XeoTo and \XeoTs have sub-MeV thresholds. Reaction thresholds are calculated using mass values in~\cite{Wang_2012}. \label{tab:isos}}
\begin{ruledtabular}
\begin{tabular}{ c c c }
Isotope & Abundance (\%) & $\nu_e$ Capture Threshold (MeV) \\
\hline 
\Tstrut$^{124}$Xe & 0.095 & 5.93 \\
$^{126}$Xe & 0.089 & 4.80 \\
$^{128}$Xe & 1.9   & 3.93 \\
$^{129}$Xe & 26.4  & 1.20 \\
$^{130}$Xe & 4.1   & 2.98 \\
$^{131}$Xe & 21.2  & 0.355 \\
$^{132}$Xe & 26.9  & 2.12 \\
$^{134}$Xe & 10.4  & 1.23 \\
$^{136}$Xe & 8.9   & 0.0903 \\
\bottomrule \bottomrule
\end{tabular}
\end{ruledtabular}
\end{table}

We first perform new calculations of the cross section for neutrino scattering into specific excited states of the product Cs nuclei using a nuclear shell model benchmarked with experimental data. The shell model is then used to predict the decay schemes of the populated excited states in \CsoTs, for which there is insufficient nuclear data. We then calculate the efficiency for tagging the de-excitation of the Cs via two techniques. First, we explore an event-topology-based analysis in which one searches for spatial separation between the CC scattering vertex and the energy deposition(s) of the emitted $\gamma$-ray(s). This technique is studied using a simulation of the scattering and decay in both target isotopes. Second, we explore a delayed-coincidence analysis in the case of \XeoTs, for which the shell model calculation predicts the existence of relatively long-lived states of the Cs daughter. We report detection efficiencies and study possible background contributions for both analyses. We close with a discussion on CNO detection in future detectors and give an additional possible application: detecting the line shift of $^{7}$Be neutrinos induced by temperature effects in the solar interior.

\section{\label{sec:model}CROSS SECTIONS AND DECAY SCHEME CALCULATIONS}

In this work we model the structure of the initial and final nuclei in the neutrino CC scattering process using the shell-model code NuShellX@MSU~\cite{Brown:2014}. The single-particle space consists of the $0g_{7/2}$, $1d_{5/2}$,
$0h_{11/2}$, $2s_{1/2}$, and $1d_{3/2}$ orbitals. We use the SN100PN effective interaction~\cite{Brown:2005}, and the single-particle energies for proton orbitals are 0.8072, 1.5623, 3.3160, 3.2238, and 3.6051~MeV, respectively, and for neutron orbitals -10.6089, -10.2893, -8.7167, -8.6944, and -8.8152~MeV.

For \XeoTs and \CsoTs we perform the calculation in the full valence space, but for \XeoTo and \CsoTo the neutron valence space needs to be truncated to make the computation time feasible. For \XeoTo and \CsoTo we use the shell-model calculation of Kostensalo \cite{Kostensalo:2020}, where the allowed configurations were restricted by setting the $\nu 0g_{7/2}$ orbital to always having the full 8 neutrons. For examining the low-lying states of our target nuclei, this approximation should be valid as this orbital has the lowest single-particle energy and the neutron numbers of the examined nuclei are close to the $N=82$ shell closure.

\begin{figure}[th]
\center
\includegraphics[width=\columnwidth]{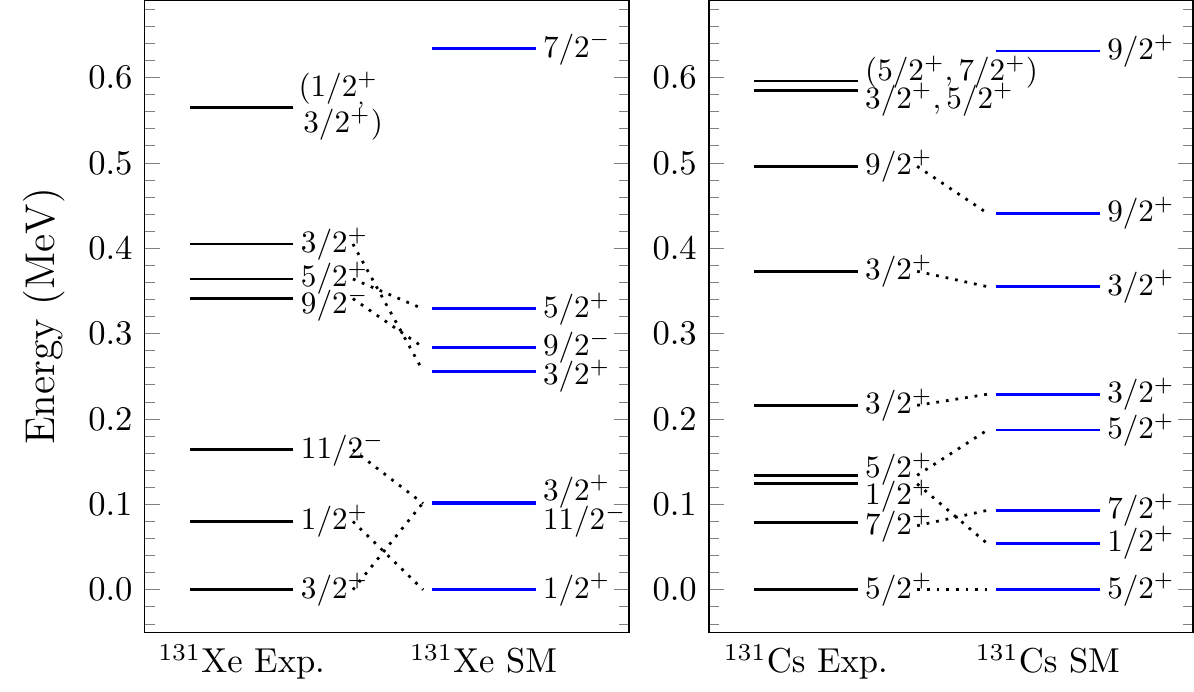}
\caption{The experimental (Exp.) and shell-model computed (SM) energy spectra of \XeoTo and \CsoTo. The dotted lines connect probable counterparts in the experimentally observed and computed spectra.}
\label{fig:levels131}
\end{figure}

The computed low-energy spectra for \XeoTo and \CsoTo are shown in Fig.~\ref{fig:levels131}, along with the experimentally determined level schemes~\cite{Khazov:2006}. The shell-model computed states correspond well with the experimental spectrum in \CsoTo. In \XeoTo the order of the first $3/2^+$ and $1/2^+$ states is not reproduced by the shell model. We use the wave function of the first $3/2^+$ state as our initial state in the present calculations of neutrino capture cross sections. 

\begin{figure}[h]
\center
\includegraphics[width=\columnwidth]{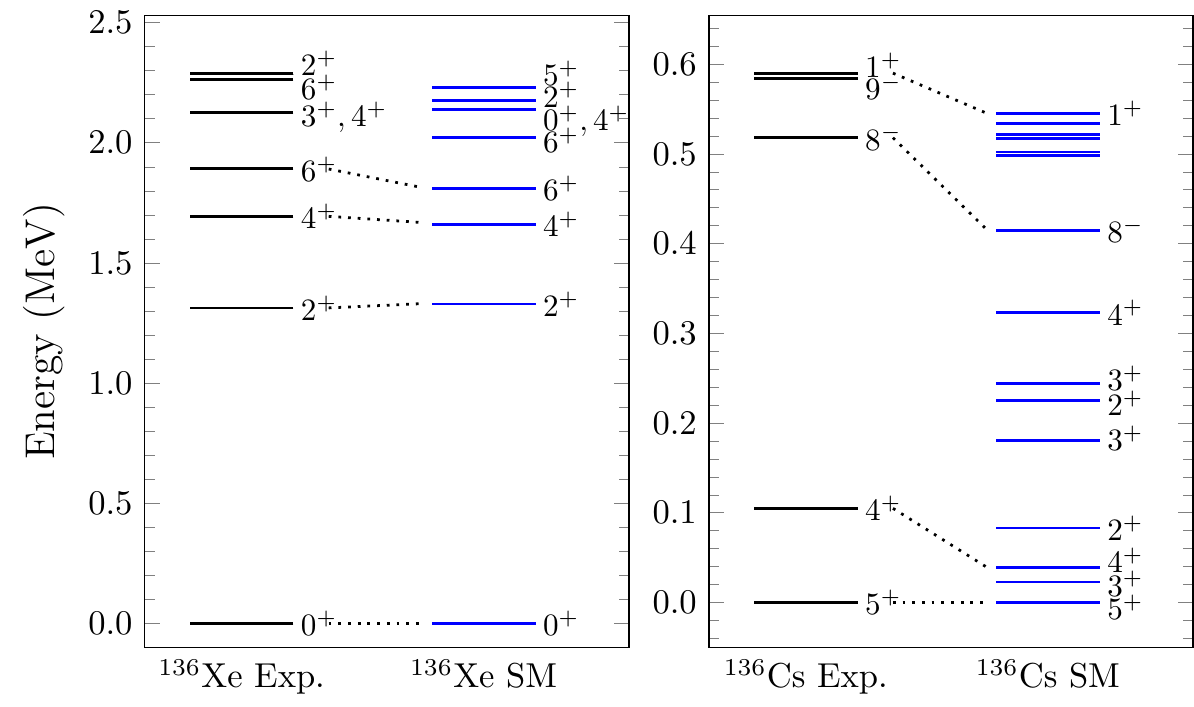}
\caption{The experimental (Exp.) and shell-model computed (SM) energy spectrum of \XeoTs and \CsoTs. The dotted lines connect probable counterparts in the experimentally observed and computed spectra.}
\label{fig:levels136}
\end{figure}

Similarly, the energy spectra of \XeoTs and \CsoTs are given in Fig.~\ref{fig:levels136}. The computed \XeoTs energy spectrum is very similar to the experimental spectrum~\cite{McCutchan:2018}. The experimental spectrum of \CsoTs is not well known and our shell-model calculation predicts several $2^+$ and $3^+$ states that are missing from the measured spectrum. The shell-model calculation appears to slightly underestimate the energy of experimentally measured states. Our calculation uses the same model space and interaction as the calculation in Ref.~\cite{Wimmer:2011}.

\subsection{\label{ssec:xsecs}Cross sections}

We compute the CC neutrino-nucleus scattering cross section to an excited final state of energy $E_x$ using the equation~\cite{Kolbe:2003,Ydrefors:2012a}
\begin{equation}\label{eqn:csfull}
\begin{split}
\frac{d \sigma_{E_x}}{d\Omega }  = & \frac{G_\mathrm{F}^2  \cos^2 \theta_\mathrm{C} \left| \mathbf{k^\prime} \right| E_{k^\prime}}{\pi (2J_i+1)} F(\pm Z_f,E_\mathbf{k^\prime}) \\ & \times \left( \sum_{J\geq 0} \sigma_\mathrm{CL}^J + \sum_{J\geq 1} \sigma_\mathrm{T}^J \right),
\end{split}
\end{equation}
where $G_\mathrm{F}$ is the Fermi coupling constant, $F(\pm Z,E_\mathbf{k^\prime})$ is a Fermi function, $\theta_\mathrm{C}$ the Cabibbo angle, and $\mathbf{k^\prime}$ and $E_\mathbf{k^\prime}$ are the three-momentum and energy of the final-state lepton, respectively. The '-' sign is chosen for neutrino scattering while the '+' sign is for antineutrino scattering. In Eq.~\eqref{eqn:csfull} $\sigma_\mathrm{CL}^J$ and  $\sigma_\mathrm{T}^J$ refer to the Coulomb-longitudinal and transverse contributions to the cross section. The details of these terms are discussed in~\cite{Pirinen:2019men} and the references therein, and the present calculations follow the same formalism discussed there. When calculating cross sections in this work, we use the experimentally measured final-state energies where available. 

The main contributions to the CC scattering cross section come from multipoles with $\Delta J = 0,1$. In the limit of zero momentum transfer the relevant operators for these multipoles simplify to Fermi and Gamow-Teller operators, and the total cross section to an individual final state reduces to~\cite{Balasi:2015}
\begin{equation}\label{eqn:csapprox}
\begin{split}
\sigma_{E_x}(E_\nu) = & \frac{G_F^2 \cos^2 \theta_\mathrm{C}}{\pi} \\  & \times \Big[ |\mathbf{k^\prime}|E_{k^\prime}F(\pm Z_f,E_\mathbf{k^\prime}) \left(B(\mathrm{F})+B(\mathrm{GT}^\pm\right)\Big],
\end{split}
\end{equation}
where
\begin{equation}\label{eqn:bf}
B(\mathrm{F}) = \frac{g_\mathrm{V}^2}{2J_i + 1} |\mathcal{M}(\mathrm{F})|^2
\end{equation}
and 
\begin{equation}\label{eqn:bgt}
B(\mathrm{GT^\pm}) = \frac{g_\mathrm{A}^2}{2J_i + 1} |\mathcal{M}(\mathrm{GT^\pm})|^2
\end{equation}
are the Fermi and Gamow-Teller reduced transition probabilities, respectively. Here $g_\mathrm{V}$ and $g_\mathrm{A}$ are the vector and axial-vector coupling constants, respectively, and $\mathcal{M}(\mathrm{F})$ ($\mathcal{M}(\mathrm{GT}$)) is the Fermi (Gamow-Teller) nuclear matrix element. 

The cross sections in this work are computed using the exact Eq.~\eqref{eqn:csfull}, accounting also for non-zero momentum transfer. However, in light of Eq.~\eqref{eqn:csapprox}, we fit our computed $B(\mathrm{GT}^-)$ values to experimentally observed values where available by applying a quenching factor. The $|\mathcal{M}(\mathrm{GT}^-)|^2$ distribution for the \XeoTs to \CsoTs process is measured directly via ($^3$He, t) charge exchange reactions~\cite{Puppe:2011zz,Frekers:2013}. For the \XeoTo to \CsoTo process we deduce the $B(\mathrm{GT}^-)$ value for the ground-state to ground-state transition from the electron-capture decay $\log ft$ obtained using the experimentally observed half life from Ref.~\cite{Khazov:2006} and phase-space factor from Ref.~\cite{Gove:1971tek}. It should be noted that Fermi transitions can only occur when  $J_i = J_f$. No experimental data is available for Fermi transitions relevant to this work, and we use the bare value of $g_\mathrm{V} = 1.0$. 

The $|\mathcal{M}(\mathrm{GT}^-)|^2$ values reported in~\cite{Frekers:2013} are determined via a charge-exchange reaction which is a strong interaction probe, allowing access to $|\mathcal{M}_\mathrm{GT^-}|^2$ directly \cite{Ejiri2019}. Thus the definition of $B(\mathrm{GT}^-)$ in Ref.~\cite{Frekers:2013} differs from our Eq.~\eqref{eqn:bgt} by the factor of $g_\mathrm{A}^2$. To bring the experimental $|\mathcal{M}(\mathrm{GT}^-)|^2$ values into the weak-interaction picture we multiply them with the bare value of $g_\mathrm{A}^2 = 1.267^2$ to get the experimental weak interaction $B(\mathrm{GT^-})$. 

\begin{figure}[th]
\centering
\includegraphics[width=\columnwidth]{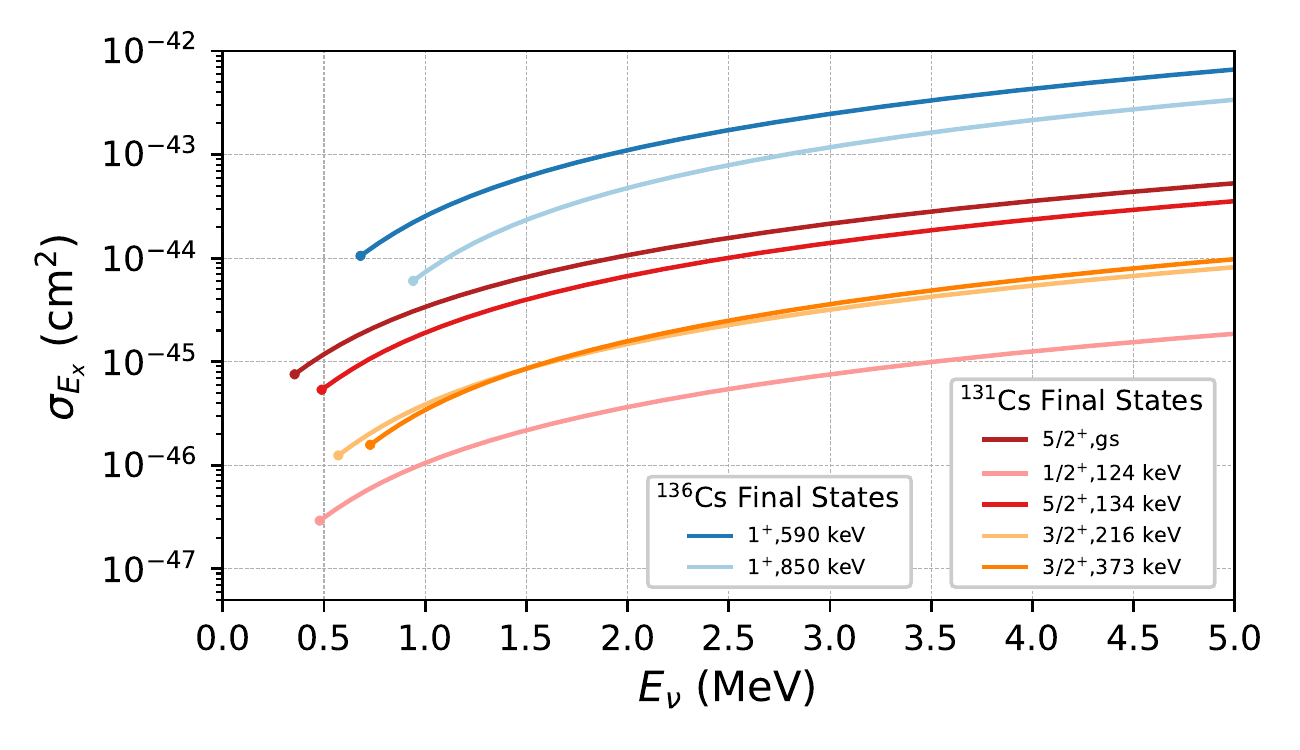}
\caption{Electron neutrino CC scattering cross sections on \XeoTo and \XeoTs to individual excited states and the ground state (gs) of the resulting Cs nuclei. The reaction threshold for scattering to each state is indicated with a marker at the beginning of each curve. }
\label{fig:raw_cross_sections}
\end{figure}

An enhanced value of $\gaeff = 1.40$ is needed to reproduce the experimentally known ground-state to ground-state $B(\mathrm{GT}^-)$ value of the \XeoTo to \CsoTo process. Due to a lack of experimental data we also use this value to compute cross sections for scattering to the excited states of \CsoTo. For the transition to the first $1^+$ state in \CsoTs we adopt a strongly quenched value of $\gaeff = 0.64$ and for the second $1^+$ state $\gaeff = 1.35$. Strong quenching of $g_\mathrm{A}$ in similar transitions between the $0^+$ ground state of an even-even nucleus and the first $1^+$ state in the neighboring odd-odd nucleus is an actively studied phenomenon, see Ref.~\cite{Suhonen:2017a} for a review. The computed cross sections are shown in Fig.~\ref{fig:raw_cross_sections}.

\subsection{Raw solar neutrino scattering rates}

The solar neutrino capture rates are obtained using the computed cross section to each of the Cs product excited states $\sigma_{E_{x}}(E_{\nu})$, folded with the differential spectrum for each solar neutrino species $\mathrm{d}\Phi_i/\mathrm{d}E_{\nu}$, and the oscillation-induced survival probability for electron-type neutrinos $P_{ee}(E_{\nu})$:
\begin{equation}
\label{eqn:avg_lee}
R_i =  \frac{ N_A}{M} \int^{\infty}_{Q+E_{x}} \sigma_{E_{x}}(E_{\nu}) \,\,P_{ee}(E_{\nu})\, \left(\frac{\mathrm{d}\Phi}{\mathrm{d}E_{\nu}}\right)_i\mathop{}\!\mathrm{d}E_{\nu} \,  ,
\end{equation}
where $Q$ is the reaction threshold, $i$ is an index representing different solar neutrino production reactions, $M$ is the xenon isotope molar mass, and $N_A$ is the Avogadro constant. We use solar neutrino flux values from the high-metallicity GS98-SFII model~\cite{Haxton:2012}. For the electron neutrino survival probability, we use the central value of the MSW-LMA solution shown in Ref.~\cite{Tanabashi:2018oca}.

Tables~\ref{tab:rate_131xe} and~\ref{tab:rate_136xe} show the event rates for solar neutrino capture on \XeoTo and \XeoTs, respectively, and compare our results with previous calculations in the literature. In the case of scattering on \XeoTo we find some disagreement between our calculations and those in Ref.~\cite{georgadze:1997zv}, which can be traced to our inclusion of neutrino oscillations, which lead to a suppression in event rates compared to those in the reference. In the case of scattering on \XeoTs, our results are within $\sim20\%$ of the calculations from Ref.~\cite{Ejiri:2013jda}, indicating good agreement. The deviation can be traced to differences in the assumed neutrino fluxes and survival probabilities, and the use of the exact Eq.~\eqref{eqn:csfull} over Eq.~\eqref{eqn:csapprox}.

\begin{table}[hb]
\centering
\caption{Raw event rates for solar neutrino CC scattering on $^{131}\textrm{Xe}$ to the ground state (gs) and to specific excited states of $^{131}\textrm{Cs}$. Also shown are the results from Ref.~\cite{georgadze:1997zv} which do not include a $\nu_e$ survival fraction. Rates are given in events per tonne of \XeoTo per year. The total $^{8}$B rate given in this work does not include scattering to higher energy states or resonances. \label{tab:rate_131xe}}
\begin{ruledtabular}
\begin{tabular}{ l l c c c c c c } 
Level $E_x$ & $J^{\pi}$ & $pp$ & $pep$ & $^{7}\textrm{Be}$ & $^{8}\textrm{B}$ & CNO & Total \\
\hline
\Tstrut gs & $5/2^+$ & 0.70 & 0.06 & 0.94 & 0.03 & 0.11 & 1.84 \\
124~keV & $1/2^+$ & - & 0.002 & 0.03 & 0.001 & 0.003 & 0.03 \\
134~keV & $5/2^+$ & - & 0.04 & 0.49  & 0.02 & 0.06 & 0.61 \\
216~keV & $3/2^+$ & - & 0.01 & 0.10 & 0.004 & 0.01 & 0.12 \\
373~keV & $3/2^+$ & - & 0.01 & 0.08 & 0.01 & 0.01 & 0.10 \\
\hline
\multicolumn{2}{l}{Total} & 0.70 & 0.12 & 1.6 & 0.06 & 0.19 & 2.7 \Tstrut\Bstrut\\
\hline
\multicolumn{2}{l}{Ref.~\cite{georgadze:1997zv} gs only} \Tstrut & 1.3 & 0.13 & 2.0 & 0.073 & 0.35 & 3.8 \\
\multicolumn{2}{l}{Ref.~\cite{georgadze:1997zv} gs+excited}  & 1.4 & 0.23 & 2.6 & 1.8 & 0.49 & 6.6 \\
\bottomrule \bottomrule
\end{tabular}
\end{ruledtabular}
\end{table}

\begin{table}[b!]
\centering
\caption{Same as Table~\ref{tab:rate_131xe} but for scattering on $^{136}\textrm{Xe}$. For comparison the results of Ref.~\cite{Ejiri:2013jda} are also shown. Rates are given in events per tonne of \XeoTs per year. Scattering to the \CsoTs ground state is highly suppressed. The total $^{8}$B rate given here does not include scattering to higher energy states or resonances. \label{tab:rate_136xe}}
\begin{ruledtabular}
\begin{tabular}{ l l c c c c c c } 
Level $E_x$ & $J^{\pi}$ & $pp$ & $pep$ & $^{7}\textrm{Be}$ & $^{8}\textrm{B}$ & CNO & Total \\
\hline
\Tstrut gs & $5^+$ & - & - & - & - & - & $\sim10^{-10}$ \\
590~keV & $1^+$ & - & 0.57 & 5.9 & 0.35 & 0.70 & 7.6\\
850~keV & $1^+$ & - & 0.21 & 0   & 0.18 & 0.18 & 0.57\\
\hline
Total & & 0 & 0.78 & 5.9 & 0.53 & 0.88 & 8.2 \Tstrut\Bstrut\\
\hline
\multicolumn{2}{l}{Ref.~\cite{Ejiri:2013jda} gs+excited} \Tstrut & 0 & 0.8 & 7.1 & 1.4 & 1.0 & 10.3 \\
\end{tabular}
\end{ruledtabular}
\end{table}

\subsection{Excited state decay scheme of \CsoTs}

\begin{figure*}[htbp]
\center
\includegraphics[width=\textwidth]{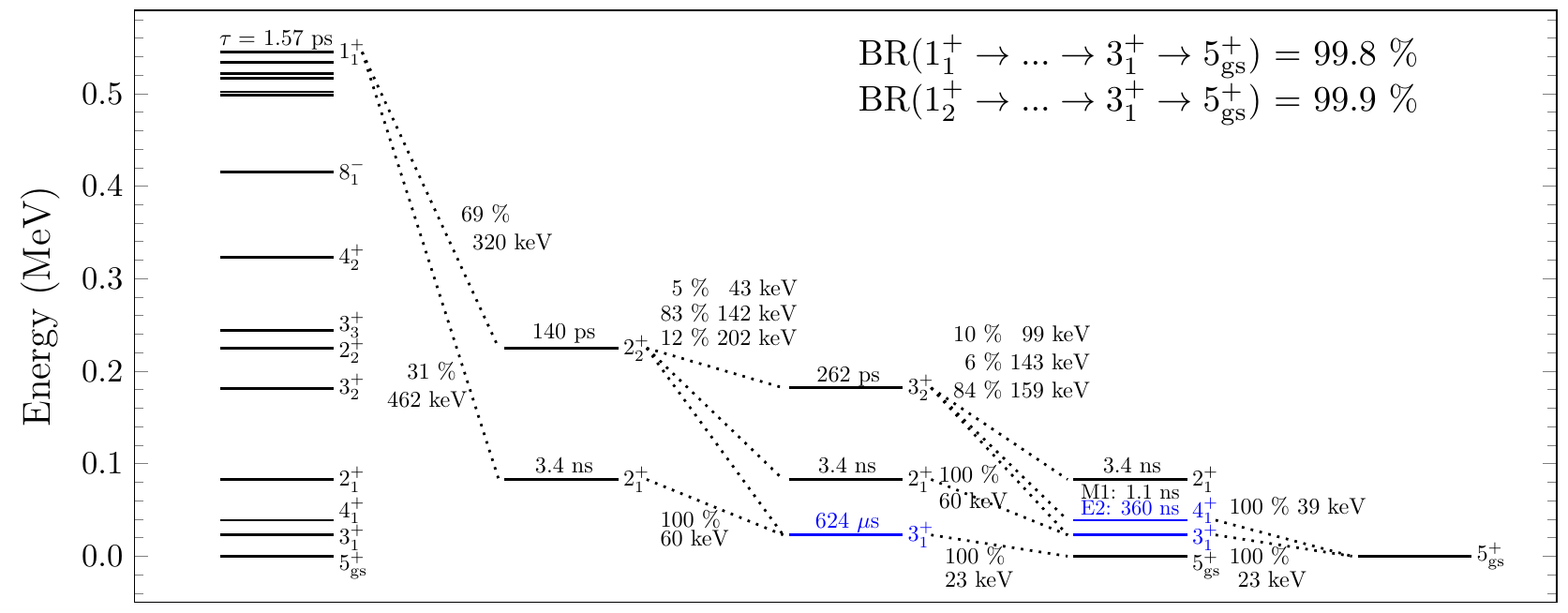}
\caption{The computed decay scheme of the first $1^+$ state of \CsoTs. The dotted lines show the possible decay branches of each state. The mean lifetime of each state is shown along with the branching ratios and energies of each transition. The standard effective charges $e_\mathrm{p} = 1.5$ and $e_\mathrm{n} = 0.5$ and bare $g$-factors were used in the calculations. }
\label{fig:gamma136}
\end{figure*}

While the level structure of \CsoTo is well measured, existing nuclear data for \CsoTs is sparse. Here we use the shell model to predict the structure and decay of low-lying states in \CsoTs which will influence the neutrino capture event signature. Fig.~\ref{fig:gamma136} shows the computed decay scheme of the first $1^+$ state of \CsoTs to the ground state, showing branching ratios for each path and lifetimes of the intermediate states. The calculation predicts that in 99.8\% of cases the first $1^+$ state rapidly decays to a very low-lying $3^+$ state at 23~keV in two or three M1-dominated steps. The decay of the 23~keV $3^+$ state is then slower, with a predicted lifetime of 624~$\mu$s, due to the low energy of the state and the E2 nature of the decay. Such a long-lived state would produce a delayed coincidence signature in modern experiments, and is discussed further in Sec.~\ref{ssec:dceff}. This finding is consistent with the propensity for $\beta\beta$-decay isotopes to exhibit low neutrino capture thresholds and for their product nuclei to have low-lying, longer-lived E1 or E2 transitions, as discussed in~\cite{Raghavan:1997ad}.

The second $1^+$ state of \CsoTs follows a very similar predicted decay path as the first. It has a predicted lifetime of 0.58~ps and decays to the two lowest-lying $2^+$ states with a branching ratio of 25\% to $2^+_2$ and 72\% to $2^+_1$. The remaining 3\% goes directly to the $3^+_1$ state via an E2 transition. The total fraction of $1^+_2$ state decays that go through the delayed $3^+_1$-to-ground state transition is 99.9\%.

The predicted lifetime of the lowest $3^+$ state depends strongly on its energy. The mean lifetime of an E2 transition is proportional to $E_x^{-5}$. By altering the state's energy while assuming that the wave function remains the same, we can obtain an estimate of the uncertainty in the predicted lifetime. Lower and upper bounds on the state energy of 10 and 100~keV lead to lifetimes of 40~ms and 0.4~$\mu$s, respectively.

The decay scheme for \CsoTs relies on the ordering of the first $3^+$ and $4^+$ states. There is an experimentally observed $4^+$ state at 105~keV, while the shell model predicts the first $4^+$ state at 39~keV. If there is no $3^+$ state between the first $4^+$ state and the $5^+$ ground state, there would likely be a fast decay path of $\Delta J = 0,1$ transitions from the first $1^+$ to the $5^+$ ground state. In our shell-model calculation the transition from the 105~keV $4^+$ state to the $5^+$ ground state is predicted to be dominantly M1 with a lifetime of $\mathcal{O}$(1~ns). However, this transition has been experimentally characterized as an E2 transition~\cite{Wimmer:2011}. If we use only the computed E2 transition strength the lifetime increases to 52~$\mu$s (assuming the predicted level energy $E_x = 39$~keV) or 360~ns (assuming the measured level energy $E_x = 105$~keV). We therefore predict that the final step in the decay scheme will have a roughly microsecond-scale (or longer) delay regardless of whether the lowest-lying excited state is a $3^+$ or $4^+$ state.

To check the accuracy of our model we compare the computed E2 transition strengths in \XeoTs to experimental values in Table~\ref{tab:136bem}. The measured values are reasonably well reproduced by the shell model calculation. However we note that the odd-odd \CsoTs is a more complex nucleus and the gamma transitions there may not be as accurate. We performed an additional cross-check by computing the known $5^+_1 \rightarrow 4^+_\mathrm{gs}$ transition of the neighboring $^{134}$Cs nucleus in the full shell-model valence space. Here the experimental energy difference is 11~keV while in the shell model it is 67~keV. The experimental $B(\mathrm{M1})$ value of the transition is $0.00363 \pm 0.00013$~W.u. and the shell-model computed value is 0.029~W.u. For the E2 transition the experimental $B(\mathrm{E2})$ value is $4.6\pm 0.6$~W.u. and our computed value is 6.4 W.u. The transition probabilities are reproduced within an order of magnitude in the shell-model calculation, and we expect similar accuracy from our calculations for \CsoTs. We do not expect the uncertainty in transition probabilities to significantly affect the total branching ratio to go through the lowest $3^+$ state.

\begin{table}[t!]
\center
\caption{Experimental (column 5) and shell-model computed (column 6) E2 strengths for $^{136}$Xe. The standard effective charges $e_\mathrm{p} = 1.5$ and $e_\mathrm{n} = 0.5$ were used in the calculations.  }
\label{tab:136bem}
\begin{ruledtabular}
\begin{tabular}{llllll}
\multirow{2}{*}{$J^\pi_{k,\mathrm{i}}$} & \multirow{2}{*}{$J^\pi_{k,\mathrm{f}}$} &$E_{\gamma,\mathrm{Exp.}}$ &$E_{\gamma,\mathrm{SM}}$ & $B(\mathrm{E2})_\mathrm{Exp.}$ & $B(\mathrm{E2})_\mathrm{SM}$ \\
 &   &  (keV) & (keV) & (W.u.) & (W.u.)  \\
\hline
\Tstrut$2^+_1$ & $0^+_\mathrm{gs}$   & 1313 & 1329 & $9.7 \pm 0.4$ & 7.5\\
$4^+_1$ &  $2^+_1$   & 381 & 331 & $1.281 \pm 0.017$ & 0.90\\
$6^+_1$ &   $4^+_1$   &  197 & 149 & $0.0132\pm0.0008$ & 0.089\\ 
$6^+_2$ & $4^+_1$ & 567 & 362 & $>0.26$ & 0.32\\
\end{tabular}
\end{ruledtabular}
\end{table}

The potential for a delayed coincidence signature in the decay of $1^+$ states in \CsoTs likely deserves experimental study for the benefit of future LXe experiments. A suitable set of experimental conditions may readily detect this signature. For example, use of the $(^{3}\textrm{He},t)$ or $(p,n)$ reactions in conjunction with an array of $\gamma$-ray and IC electron detectors could verify the predictions reported here.

\section{\label{sec:event_signatures}EVENT SIGNATURES IN LIQUID XENON DETECTORS}

Interactions in LXe TPCs are measured via two channels: scintillation light, which is released promptly from excited Xe$_2^*$ dimers at the interaction vertex; and the ionized charge, which is drifted to a collection plane by an electric field applied across the target volume. The scintillation signal develops over a timescale of $\mathcal{O}(100)$~ns, which is dominated by the intrinsic lifetime of the singlet and triplet dimer states (4~ns and 24~ns, respectively), the timescale of electron-ion recombination (up to 10's of ns, depending on the applied electric field~\cite{Hogenbirk_2018}), and the optical transport time for photons in the detector~\cite{Akerib:2018kjf}. The latter is expected to be several 10's of nanoseconds in next-generation detectors. The charge signal drifts across the detector with a velocity of $\sim$1--2~mm/$\mu$s, and is collected over a timescale of several microseconds. Detectors with multi-tonne targets (with linear dimensions $\mathcal{O}(1)$~m) therefore have event windows of up to a few milliseconds. Ions which are created in the active target region drift with comparatively slower speeds on the order of $10^{-6}$~mm/$\mu$s~\cite{Albert:2015vma}, which is negligible on the timescale of a single event in the detector.

The light and charge signals provide energy and position information about each interaction in the detector. The energy can be reconstructed from the magnitude of the light and charge signals, either individually or using a linear combination of the two. The linear combination provides the best energy resolution due to anticorrelation between the light and charge channels~\cite{ContiAnticorrelation}. The $x$- and $y$-coordinates of the interaction can be measured by the position of the charge signal on the collection plane, while the $z$-coordinate is measured by the time difference between the prompt scintillation signal and the drifted charge signal. This results in position resolution which is typically of $\mathcal{O}(1)$~cm or better. Events with multiple interactions, e.g. a $\gamma$-ray which Compton scatters before being fully absorbed, will create multiple interaction vertices in a single event. In this case, the scintillation signals from all interactions will be recorded simultaneously and merged, but the individual charge signals may be separated if the distance between them is greater than the detector's resolution in $x/y$ (dictated by the pitch/pixelization of the readout plane) and/or $z$ (dictated by the time resolution for the charge signals).

The combination of fast timing information, few-millimeter-scale position resolution for each vertex, and good energy resolution allows the use of advanced analysis techniques that can suppress backgrounds in searches for rare processes. Here we explore the possibility of tagging neutrino capture events on \XeoTo and \XeoTs that populate excited states in \CsoTo and \CsoTs using an event-topology-based analysis and a delayed coincidence analysis.

\subsection{Event topology analysis\label{ssec:topology}}

The neutrino capture events of interest are comprised of a primary, fast ($\mathcal{O}(100)$~keV typical) electron and $\gamma$-rays/IC electrons from the excited Cs de-excitation. We perform an analysis with the goal of tagging these events by resolving the emitted $\gamma$-rays which may be absorbed or Compton scatter and produce multiple interaction vertices. The summed energy of these vertices must recover the excited state energy. The analysis presented here does not use the temporal information of each event and is therefore meant to be distinct from the delayed coincidence method discussed in Sec.~\ref{ssec:dceff}.

Full particle-tracking simulations of neutrino capture events were performed using the \textsc{GEANT4}-based~\cite{Agostinelli:2002hh} XeSim package~\cite{Althueser2019}. The event generator models neutrino signal events by generating two primary particles: an electron with energy $E_{\nu} - Q - E_x$ and a Cs nucleus in an excited state. The incident neutrino energy can be drawn from any component of the solar neutrino spectrum. The events are simulated in a cylinder of LXe that is 1.2~m in both diameter and height, which approximately models few-tonne detectors such as nEXO, LZ, or XENONnT. The de-excitation of \CsoTo is simulated using the nuclear levels, branching ratios, and lifetimes from the appropriate data file in the built-in \textsc{PhotonEvaporation3.2} library. For \CsoTs, we replace the existing data file with one containing the decay scheme shown in Fig.~\ref{fig:gamma136}. 

To mimic the response of an actual detector, simulated neutrino events are processed by the algorithm developed in Ref.~\cite{Wittweg:2020fak}, which groups individual energy deposits into ``clusters" within the assumed position resolution of the detector. The energy for each cluster is also smeared using a Gaussian with width $\sigma_E$ defined by
\begin{equation}
    \frac{\sigma_E}{E} = \frac{a}{\sqrt{E}} + b ,
    \label{eqn:energy_resolution}
\end{equation}
where $E$ is the total energy deposited in the cluster. We use the $a$ and $b$ parameter values from the charge-plus-light and charge-only energy resolution fits in Refs.~\cite{Aprile:2020yad} and \cite{XENON100_DetPaper}, respectively. The values are $a=0.317 \, \textrm{keV}^{1/2}$ and $b=0.015$ for the combined charge-plus-light energy and $a = 0.75 \, \textrm{keV}^{1/2}$, $b = 0.035$ for charge-only. In a real experiment, charge detection of low-energy clusters may be limited by a threshold, leading to a loss of some clusters in an event. We omit this effect here for simplicity. The detection efficiency of each Cs state is evaluated by checking if, for the $N$ clusters in the event, there is any sub-combination of $N-1$ clusters which fall within $2\sigma$ of the state energy (the $N^{\textrm{th}}$ cluster is assumed to be the signal from the primary fast electron). The efficiency is approximately independent of incoming neutrino energy as the primary fast electron preferentially forms a single-site energy cluster. 

The calculated efficiencies for $^{7}$Be neutrino scattering to each of the low-lying states in \CsoTo and \CsoTs are shown in Table~\ref{tab:det_eff} assuming a 5~mm resolution in $z$ and either 10~mm or 30~mm resolution in $x-y$. An efficiency of $\sim$50\% is found for the most favorable states. The position-based tagging efficiency is highly dependent on the energy of the state: the higher-energy states are likely to emit more energetic photons, which have longer mean free path, while lower-energy states are more likely to emit lower-energy photons and IC electrons, which deposit energy very close to the primary nucleus. The results presented here cover the range of expected position resolutions in next-generation detectors: dark matter experiments quote projections as good as 3~mm in the $z$ dimension~\cite{LZ_0nuBB_sensitivity2020} and 15~mm in the $x-y$ dimension~\cite{Agostini:2020adk}. The efficiencies reported here do not change drastically with $x-y$ resolution due to the relatively good resolution assumed for $z$. Worse resolution in $z$ would require better $x-y$ resolution to maintain the same tagging efficiency.

\begin{table}[t]
\centering
\caption{Detection efficiencies for $^{7}$Be neutrino capture events leading to the given excited states of \CsoTo and \CsoTs obtained using the event-topology analysis. Results are shown using a fixed resolution in $z$ of 5~mm and two values of the $x-y$ resolution to give a reasonable range of experimental possibilities. The relaxation process in \CsoTs is assumed to follow the decay scheme in Sec.~\ref{sec:model}. \label{tab:det_eff}}
\begin{ruledtabular}
\begin{tabular}{ l l c c } 
 &  & \multicolumn{2}{c}{$x-y$ Resolution} \\
Isotope & State & 30~mm (\%) & 10~mm (\%) \\ \hline 
\Tstrut\multirow{4}{*}{$^{131}\textrm{Cs}$} & $124$~keV & 3.6 & 4.3 \\
 &  $134$~keV & 4.3 & 5.4 \\
 &  $216$~keV & 32.4 & 39.1 \\
 &  $373$~keV & 47.5 & 51.1 \\
 \hline
\Tstrut\multirow{2}{*}{$^{136}\textrm{Cs}$} & $590$~keV & 43.9 & 46.6 \\
 &  $850$~keV & 49.1 & 54.8 \\
\end{tabular}
\end{ruledtabular}
\end{table}

\subsection{\label{ssec:dceff}Delayed coincidence analysis}

In order for two events to be distinguishable in a coincidence analysis, their primary scintillation signals must be separated in time by more than the typical reconstructed pulse width. This quantity depends on multiple factors including the xenon scintillation decay time, the size of the detector, the chosen photosensor coverage and characteristics, and the detector electronics. Here we take $t_1$ to represent the minimum time separation between the first and second scintillation signals for which these two pulses are distinguishable.

\begin{figure}[t]
\centering
\includegraphics[width=\columnwidth]{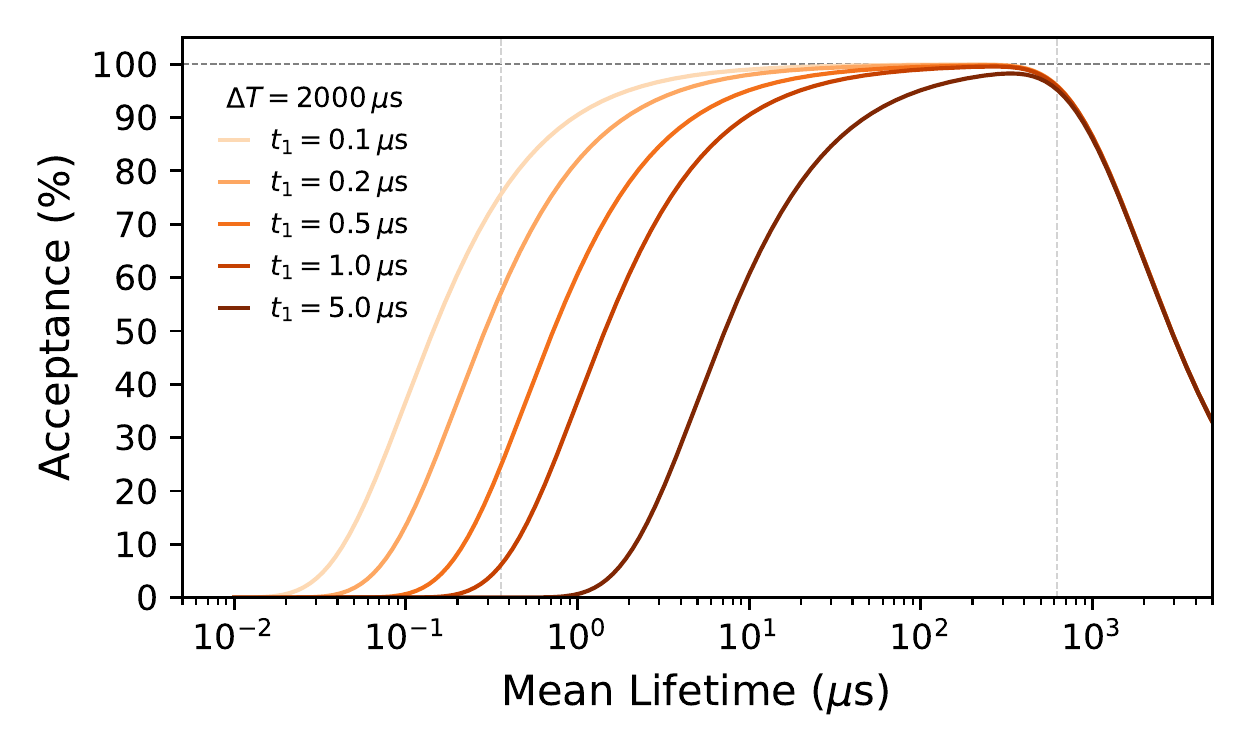}
\caption{Acceptance for delayed signals in a coincidence window $\Delta T=2000$~$\mu$s as a function of the mean lifetime of the nuclear state calculated using Eq.~\eqref{eqn:dc_acc}. Shown are curves for several choices of the coincidence window start time $t_1$. Vertical lines show the predicted 360~ns and 624~$\mu$s lifetimes of the $4^+$ and $3^+$ states discussed in the text. \label{fig:timing_eff}}
\end{figure}

The acceptance for a delayed signal to arrive in a coincidence window of length $\Delta T$ which begins at time $t_1$ after a primary pulse is
\begin{equation}
\label{eqn:dc_acc}
\begin{split}
\epsilon(\tau,t_1,\Delta T) & =  \frac{1}{\tau} \int^{t_1+\Delta T}_{t_1} e^{-t/\tau} \mathrm{d}t \\
     & = e^{-t_1/\tau}-e^{-(t_1+\Delta T)/\tau} \, ,
\end{split}
\end{equation}
where $\tau$ is the mean lifetime of the excited nuclear state. In the present work we will take $t_1=0.5$~$\mu$s and assume that two primary scintillation pulses will be distinguishable if they are separated by at least this amount. As a coincidence window length we take $\Delta T=2000$~$\mu$s, consistent with the expected event acquisition length of a future TPC with drift length $\gtrsim1.2$~m. For the decay of the 23~keV $3^+$ state in \CsoTs with mean lifetime 624~$\mu$s, this coincidence window accepts 96\% of decays. If in reality the first $3^+$ state lies higher in energy than the shell model predicts and the coincidence results from the experimentally measured E2 transition of the 105~keV $4^+$ state with our calculated lifetime of 360~ns, the acceptance is 25\%. In this case the loss in efficiency comes from the minimum time separation requirement. Fig.~\ref{fig:timing_eff} shows more generally how the acceptance changes with the excited state lifetime for a few choices of the window start time $t_1$ and our chosen window duration. 

To reduce backgrounds further, two additional requirements are imposed to select neutrino capture events. First, the energy associated with the delayed pulse should be consistent with the energy of the final de-exciting nuclear state. For our analysis we will accept secondary pulses in a $3\sigma$ energy window centered on the excited state energy (the efficiency is then 99.7\%). Second, the reconstructed position of the delayed event vertex must be sufficiently close to that of the prompt signal, which is comprised of all energy deposits that occur before the long-lived nuclear state is reached. We require that the delayed signal be located within a 5~cm radius of the weighted-average position of the prompt energy deposits. Simulations from Sec.~\ref{ssec:topology} show that such a cut will accept 93.5\% of neutrino capture events. The total efficiencies for selecting events from the 624~$\mu$s and 360~ns states are then 89.8\% and 23.4\%, respectively.

\subsection{Signal rates in future detectors}

We now fold the calculated detection efficiencies with the expected interaction rates to predict the number of tagged events in a given experiment. We consider the cases of neutrino capture on \XeoTo and \XeoTs separately. We estimate the event rates that would be observed in DARWIN~\cite{Aalbers:2016jon}, which proposes to use a 40-tonne target of natural xenon, and nEXO~\cite{NEXO_pCDR:2018eqi}, which will deploy a $\sim$4-tonne target of xenon enriched to 90\% in \XeoTs.

The low cross sections for neutrino capture on \XeoTo result in event rates which are likely too low for next-generation experiments. The most promising signal is topologically-tagged events produced by populating the 373~keV level in \CsoTo. Given the efficiencies and interaction rates calculated above and a natural abundance of 21.2\%, we calculate an event rate of 0.3~(0.04)~events/yr for $^7$Be (CNO) neutrinos in DARWIN. For a ten-year exposure, this would produce a total of $\mathcal{O}(1)$ events, from which it would be difficult to draw any conclusions. An experiment using xenon enriched in \XeoTs will have essentially no \XeoTo in the target volume, so we expect no events from this process in nEXO. 

In the case of \XeoTs, we find more promise in the possibility of a delayed coincidence analysis, which potentially allows higher detection efficiency for neutrino capture events. In addition, the cross section for exciting the 590~keV state in \CsoTs is substantially higher than that for any of the low-lying states in \CsoTo. The expected tagged event rates for $^{7}$Be and CNO solar neutrino scattering to the 590~keV state in \CsoTs are 5.3 and 0.63 events per tonne of \XeoTs per year. DARWIN and nEXO will contain approximately 3.5 and 3.6~tonnes, respectively, meaning they will each observe $\approx19$ $^{7}$Be neutrino capture events per year and $\approx2.2$ CNO neutrino captures per year. We have neglected to include the effect of an energy threshold when calculating the detection efficiency for the state at 23~keV; while this is expected to be a good approximation for experiments detecting the charge via electroluminescence amplification such as DARWIN, a dedicated study would need to be made for experiments such as nEXO where the ionized charge will be measured directly (and may therefore be dominated by readout noise in the energy regime below $\sim$100~keV).

\section{\label{sec:backgrounds}BACKGROUND CONSIDERATIONS}

The backgrounds in modern LXe TPCs fall into two broad categories: internal and external sources. External sources arise primarily from radioactivity present in the laboratory environment and the detector construction materials. These can generally be suppressed by defining an inner fiducial volume which is shielded by the LXe near the detector edge. Internal sources arise primarily from decay of radioisotopes dissolved in the LXe and from solar neutrino scattering on xenon nuclei and electrons. In this section we focus on internal backgrounds, which cannot be removed with the use of a fiducial volume cut and must be rejected using the tagging strategies defined above.

\subsection{Backgrounds for topological analysis}

The primary source of background events for the topological analysis are decays of $^{214}$Pb. This isotope enters the LXe volume as a daughter of $^{222}$Rn which emanates from surfaces in the LXe system. Approximately 90\% of $^{214}$Pb $\beta$ decays leave the $^{214}$Bi daughter in an excited state which decays rapidly ($\tau \lesssim100$~ps) to the ground state by emitting $\gamma$-rays and IC electrons. The topology of these $\beta+\gamma/e^-$ events can mimic that of the neutrino capture signal when the incident neutrino energy falls below the $^{214}$Pb $\beta$-decay $Q$-value of 1018~keV. For higher energy neutrinos, the dominant multi-scatter background is likely from external $\gamma$-rays which may be suppressed with a detector-specific fiducial cut, and we therefore do not quantify this background here. 

The background from $^{214}$Pb is assessed by simulating its decay and processing the events in the framework described in Sec.~\ref{ssec:topology}. Table~\ref{tab:bkg_topo_eff} shows the efficiency with which $^{214}$Pb decays pass the selection used to search for each of the Cs excited states as well as the expected background rate in a detector containing 1~$\mu$Bq/kg of $^{222}$Rn. For comparison, the table also shows the same quantities for the $^{7}$Be solar neutrino capture signal. We conclude that a future experiment could only hope to detect the capture signal by this method provided the detector be practically radon-free, containing $\lesssim 10^{-5}$~$\mu$Bq/kg $^{222}$Rn.

\begin{table}[t]
\centering
\caption{Efficiency (Eff.) for background $^{214}$Pb $\beta$-decays to pass the topological analysis criteria for neutrino captures to each excited state and the resulting background rate in a detector with 1~$\mu$Bq/kg $^{222}$Rn using natural xenon. Shown for comparison are the same quantities for the capture signal from $^{7}$Be neutrinos. The assumed position resolutions are 5~mm in $z$ and 10~mm in $x-y$.} 
\label{tab:bkg_topo_eff}
\begin{ruledtabular}
\begin{tabular}{ l l c c c c }
 &  & \multicolumn{2}{c}{$^{214}$Pb Background} & \multicolumn{2}{c}{$^{7}$Be Signal} \\
 & &  & Rate  &  & Rate  \\ 
Isotope & State & Eff. (\%) & (evts/t/yr) & Eff. (\%) & (evts/t/yr) \\ \hline 
\Tstrut\multirow{4}{*}{$^{131}\textrm{Cs}$} & $124$~keV & 4.7 & 1482 & 4.3 & $2.4\times10^{-4}$ \\
 &  $134$~keV & 4.9 & 1545 & 5.4 & $5.7\times10^{-3}$ \\
 &  $216$~keV & 9.7 & 3059 & 39.1 & $8.1\times10^{-3}$\\
 &  $373$~keV & 23.3 & 7348 & 59.1 & $8.8\times10^{-3}$\\
 \hline
\multirow{2}{*}{$^{136}\textrm{Cs}$} & $590$~keV & 6.9 & 2176 & 46.6 & 0.59 \Tstrut\\
 &  $850$~keV & 2.6 & 820 & 54.8 & - \\
\end{tabular}
\end{ruledtabular}
\end{table}

\subsection{Backgrounds for delayed coincidence}
\begin{figure}[tb]
\centering
\includegraphics[width=\columnwidth]{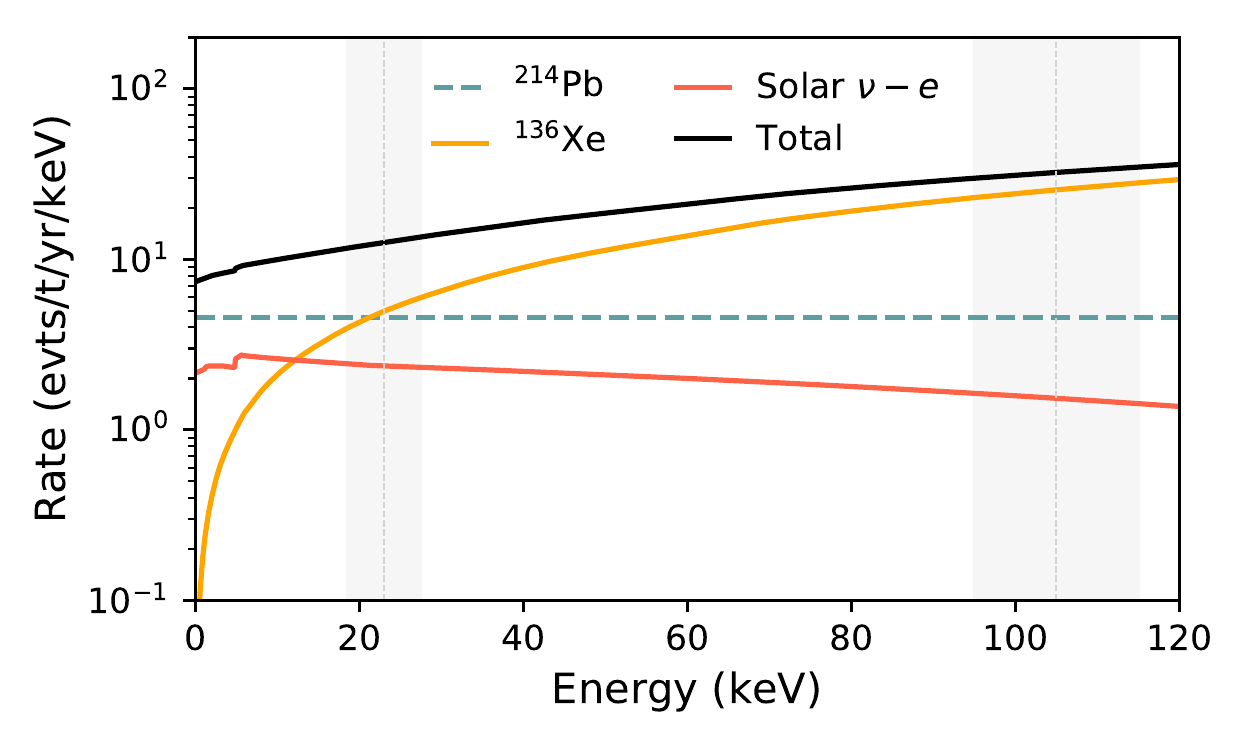}
\caption{Energy spectra of typical electron recoil backgrounds in a detector based on natural xenon. Included are the spectra from \tnbd decay of \XeoTs~\cite{Kotila:2012zza,KotilaWeb}, the scattering of $pp+^{7}\textrm{Be}+^{13}\textrm{N}$ solar neutrinos off atomic electrons (scaled according to~\cite{Chen:2016eab} below 30~keV), and $\beta$ decay of $^{214}\textrm{Pb}$, taken as flat in this energy range. The assumed $^{222}\textrm{Rn}$ concentration is 1~$\mu$Bq/kg. In a detector enriched to 90\% \XeoTs, the background from \tnbd will be larger by a factor of 10. Vertical bands locate the $\pm3\sigma$ energy regions around the low-lying $3^+$ (predicted here) and $4^+$ (measured) states of \CsoTs at 23~keV and 105~keV, respectively. \label{fig:bkgs}}
\end{figure}

The stringent selection requirements for the delayed coincidence signature defined in Sec.~\ref{ssec:dceff} lead to an estimated background rate that is very small. The expected background is comprised of accidental coincidences of low-energy events (within $3\sigma$ of the delayed energy) following any other event with energy consistent with the prompt neutrino capture signature. In this study we consider three sources of low-energy events: \tnbd decay of \XeoTs, scattering of solar neutrinos on atomic electrons, and $\beta$ decays of $^{214}$Pb. These are expected to dominate the background rate in the 0--100~keV energy range relevant for tagging the delayed signals in \CsoTs (see for example Ref.~\cite{LZ_WIMP_Sensitivity2019}). The energy spectra for these backgrounds are shown in Fig.~\ref{fig:bkgs}. We note that while a delayed timing signature from the $\beta-\alpha$ decay sequence $^{214}\textrm{Bi}\rightarrow^{214}\textrm{Po}$ ($\tau = 236$~$\mu$s) will also be present, the large (7.7~MeV) energy of the secondary $\alpha$ pulse excludes these events from our selection.

From the rates in Fig.~\ref{fig:bkgs}, we derive the expected event rate within a 5~cm radius (corresponding to a 1.47~kg mass) and $3\sigma$ energy window centered at 23~keV and 105~keV to be 0.23~evts/yr and 1.1~evts/yr, respectively. This yields probabilities for coincidence with a primary event in the 2000~$\mu$s search window of $<10^{-10}$. We note that this estimate is relatively robust against theoretical uncertainties on the energy of the delayed \CsoTs state, as the event rate changes by less than a factor of $\sim3$ across the relevant energy range. In a detector using a \XeoTs-enriched target, the background rate from \tnbd will be approximately an order of magnitude higher, leading to a correspondingly higher coincidence probability.

The total number of events in an experiment that mimic the primary neutrino capture signal then determines the overall background estimate. As an example, the nEXO experiment is expected to observe approximately $10^8$ events in the 700--3500~keV range over ten years of operation~\cite{Albert:2017hjq}. Given the coincidence rates calculated above we expect $\ll 1$ background event to pass our selection, making a delayed coincidence analysis essentially background free.

\section{\label{sec:discussion}DISCUSSION}

We conclude from the above assessment of efficiencies and projected backgrounds that a delayed coincidence analysis can provide a background-free detection of $\mathcal{O}(1)$~MeV solar neutrinos via CC capture on \XeoTs in next-generation LXe experiments. In this section we discuss two applications of this technique.

\subsection{Outlook for CNO neutrino detection}

The detection of CNO neutrinos as a means of completing a detailed picture of power generation and dynamics in our Sun has been a long-standing and sought-after experimental goal for modern solar neutrino experiments. After much experimental effort, the first detection of CNO neutrinos was recently reported by the Borexino collaboration~\cite{borexinoCNO}. Here we briefly mention the prospects for CNO detection using the capture signature on \XeoTs.

Excluding the lines from $^{7}$Be and $pep$, the flux of solar neutrinos that lies between the $pp$ endpoint at 0.42~MeV and the CNO endpoint at roughly 1.73~MeV is dominated by the CNO contribution. The ratio of CNO flux to the combined $^{8}\textrm{B}+hep$ flux in this region is approximately $4\times10^3$. Therefore, any event consistent with a capture signal on \XeoTs with reconstructed energy in this window (less the reaction threshold $Q$) is overwhelmingly likely to be a CNO event, while events around the $^{7}$Be and $pep$ energies can be easily removed from that sample. As an example, we project using Table~\ref{tab:rate_136xe} that a ten-year exposure of the 90\% enriched, 4~tonne fiducial mass of the nEXO detector would expect to see approximately 32 total CNO events with a background of $<1$ event using the delayed coincidence technique. In combination with the uncertainty on the measured $B(\mathrm{GT}^-)$ values~\cite{Frekers:2013}, the fractional uncertainty on this count rate is roughly a factor of two lower than that obtained by the current energy spectrum fit in Borexino.

It is interesting to note that this approach contrasts with the case of neutrino-electron scattering in LXe studied in Ref.~\cite{Newstead:2018muu}. There, a CNO detection at $\sim 3\sigma$ significance in a dark matter experiment could only be achieved with significant depletion of the \XeoTs isotope. By contrast, our discussion suggests that this isotope may be well-suited for such a detection with the caveat that one does not accumulate enough events to precisely reconstruct the energy spectrum. However, it has the benefit that such a search can proceed concurrently with the primary science goals of the experiment and the cost of isotope separation can be avoided.

The opportunity for CNO detection in other future experiments was explored in Ref.~\cite{Cerdeno:2017xxl}. Specifically, a CNO measurement in a five-year exposure of the SNO+~\cite{Andringa:2015tza} detector may be possible provided this run is free of the \tnbd~decay isotope $^{130}$Te. In fact, an argument could be made in favor of a search for CNO neutrino capture on the $^{130}$Te isotope itself during the SNO+ primary physics run. However, the capture rate on $^{130}$Te calculated by Ejiri in~\cite{Ejiri:2013jda} suggests that the roughly 0.8~tonnes of isotope deployed in SNO+ would not be competitive with the analogous rate in a LXe experiment as discussed here. Existing nuclear data for the product $^{130}$I does suggest a possible delayed coincidence signature~\cite{A130sheet}; however the energies of the relevant excited states fall below the current SNO+ energy threshold.

\subsection{Detecting the shift of $^{7}\textrm{Be}$ neutrinos}

In Ref.~\cite{Bahcall:1994cf}, Bahcall points out that the neutrino energy line shape resulting from the electron-capture decay of $^{7}\textrm{Be}$,
\begin{equation}
\label{eqn:7be}
    e^{-}+^{7}\textrm{Be} \rightarrow ^{7}\textrm{Li}+\nu_e \, ,
\end{equation}
in the Sun is distorted compared to that observed by a laboratory source of $^{7}\textrm{Be}$. In the stellar core, electrons are captured from unbound continuum states in the thermal bath, and so the resulting neutrino energy spectrum is broadened by thermal effects. The shape of the broadened spectrum and, in particular, its mean energy therefore depend on the Sun's central temperature distribution. The assumptions used in~\cite{Bahcall:1994cf} predict a mean neutrino energy which is 1.27~keV higher than the monoenergetic 861.8~keV line expected from Eq.~\eqref{eqn:7be} in the laboratory.

Detection of this shift requires reconstruction of the incident neutrino energy, making neutrino capture an ideal process for this measurement. The predicted capture rate on \XeoTs and the potentially background-free event signature may enable such a detection in future LXe experiments. Capture events on \XeoTs result in a a total energy deposit of $E_{\nu}-Q$, where $Q=90.3$~keV is the reaction threshold. The total event energy expected from a laboratory source of $^{7}\textrm{Be}$ neutrinos is therefore 771.5~keV.

To quantify the likelihood of observing the broadened neutrino spectrum of~\cite{Bahcall:1994cf} we compute the significance with which an experiment would reject the ``laboratory energy" hypothesis in favor of the hypothesis that the incident neutrinos follow the spectrum predicted by Bahcall. As a test statistic TS we use the distance between the mean energy of events $\bar{E}_{\mathrm{obs}}$ and the energy expected from the laboratory neutrino source $E_{\mathrm{lab}}=771.5$~keV normalized by the error on the observed mean energy:
\begin{equation}
    \mathrm{TS} = \frac{ \bar{E}_{\mathrm{obs}}-E_{\mathrm{lab}} }{ \sigma_{E_{\mathrm{obs}}}/\sqrt{N_{\mathrm{obs}}} } \, .
\end{equation}
Here, $\sigma_{E_{\mathrm{obs}}}$ is the standard deviation of observed energies and $N_{\mathrm{obs}}$ is the observed number of events in the exposure. For each assumed exposure of \XeoTs, a set of Monte Carlo experiments were performed in which a Poisson random number (with mean calculated using the rate in Table~\ref{tab:rate_136xe}) of energies are sampled from the broadened neutrino spectrum and smeared by a Gaussian detector energy resolution. The TS value from each experiment is then used to calculate a $p$-value using the distribution of TS's calculated from experiments which sample the monoenergetic laboratory spectrum.

\begin{figure}[t]
\centering
\includegraphics[width=\columnwidth]{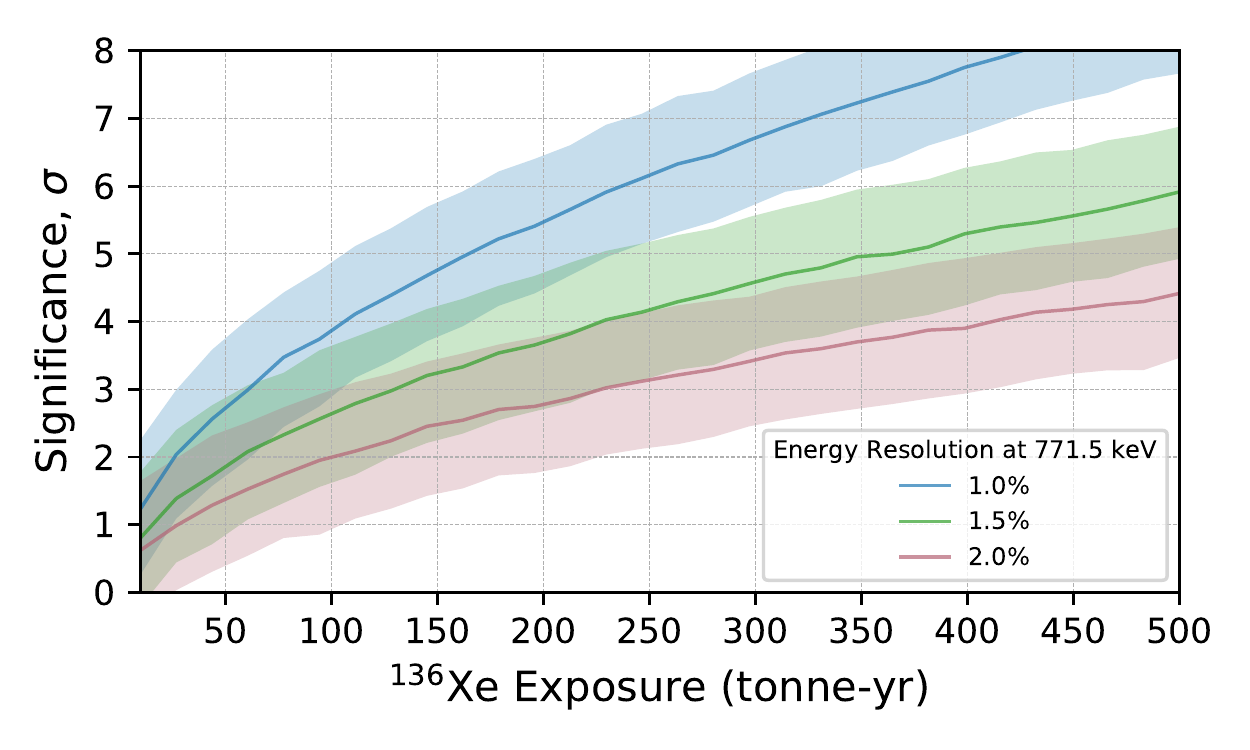}
\caption{Median and $\pm1\sigma$ discovery significance bands for detecting the broadened and shifted 862~keV $^{7}$Be neutrino line of Ref.~\cite{Bahcall:1994cf} as a function of \XeoTs exposure. Shown are three values of detector energy resolution at 771.5~keV. \label{fig:shift}}
\end{figure}

The median significance and surrounding 68\% interval for observing the broadened spectrum is shown as a function of \XeoTs exposure in Fig.~\ref{fig:shift}. In the figure we show curves for three benchmark values of the detector's multi-site energy resolution at 771.5~keV. We note that a resolution of approximately 1.5\% has been demonstrated in the XENON1T detector~\cite{Aprile:2020yad}. Our results rely on the assumption that the absolute energy scale can be known to sufficient accuracy.

Our results indicate that a detection above $3\sigma$ may be attainable, for example, with an 80~tonne natural xenon experiment that runs for ten years (achieving a \XeoTs exposure of 71~tonne-yr). We also note that a more severe shift of the neutrino energy spectrum (e.g. from a re-evaluation of the solar parameters assumed in~\cite{Bahcall:1994cf}) would reduce the exposure required for a detection.

\section{\label{sec:conclusion}CONCLUSION}
We have considered the detection of solar neutrino CC interactions on the isotopes \XeoTo and \XeoTs in LXe TPC experiments. Nuclear shell model calculations were used in conjunction with available experimental data to calculate the CC scattering cross sections and event rates to low-lying excited states in the concomitant Cs product nuclei. The shell model was further used to predict the decay scheme of $1^+$ states in \CsoTs through $3^+$ and $4^+$ states with intermediate lifetimes which would permit a delayed coincidence signature. We suggest that this prediction be investigated with a dedicated experiment.

Two techniques were investigated for identifying the $\gamma$'s/IC electrons emitted during the resulting Cs relaxation: spatial separation via an event-topology-based analysis and time separation based on delayed coincidence. The topological reconstruction yields detection efficiencies up to $\sim$50\% and can reduce backgrounds by factors of 2--4, depending on which excited state is populated. However, the background from dissolved $^{214}$Pb is likely to be several orders of magnitude higher than the neutrino scattering rate, making such a search practically impossible in a detector that is not radon-free. By contrast, a delayed coincidence analysis can reach detection efficiencies of $>$90\% and reject backgrounds by a factor of $\approx10^{10}$, making this signature a promising avenue for the detection of solar neutrinos on \XeoTs in future experiments.

We explore applications of solar neutrino measurements using the delayed coincidence technique in LXe TPCs. We find that next-generation experiments may detect the CNO neutrino flux with smaller uncertainty than that in current experiments. Finally, we determine that a next-generation experiment with suitable energy resolution may detect the broadened neutrino spectrum from $^{7}$Be decay in the Sun. Both of these measurements would provide further probes of solar models.

\section*{ACKNOWLEDGMENTS}

The authors would like to thank Joel Kostensalo for performing shell model calculations of \XeoTo and \CsoTo in the large valence space and for discussions on experimental $\log(ft)$ values. We thank Andrea Pocar, Giorgio Gratta, Henrique Ara\'{u}jo, and Kevin Lesko for comments on the manuscript. This work was supported by the U.S. Department of Energy Office of Science under contract number DE-AC02-05CH11231 and by DOE-NP grant DE-SC0017970. B.L. acknowledges the support of a Karl Van Bibber fellowship from Stanford University. This work has been partially supported by the Academy of Finland under the Academy project no. 318043.

\FloatBarrier
\bibliographystyle{apsrev4-2}
\bibliography{neutrino}

\end{document}